\documentclass[
 reprint,
 showpacs,
 amsmath,amssymb,
 aps
]{revtex4-1}
\usepackage{verbatim}
\usepackage{ulem}

\usepackage{lipsum}
\usepackage{amsmath}
\usepackage{amssymb}
\usepackage{graphicx}
\usepackage{dcolumn}
\usepackage{color}
\usepackage{xcolor}
\usepackage{endnotes}
\usepackage{bm}
\usepackage{epsfig}
\usepackage{array}
\usepackage{subfigure}
\usepackage{graphicx}
\usepackage{dcolumn}
\usepackage{bm}
\usepackage{color}
\usepackage{booktabs,multirow}
\usepackage{bigstrut}

\usepackage[colorlinks,linkcolor=red,anchorcolor=blue,citecolor=green]{hyperref}

\usepackage{rotfloat}
\usepackage{cleveref}
\usepackage{natbib}

\def \bea{\begin{eqnarray}}
\def \eea{\end{eqnarray}}

\def \ba{\begin{array}{cccc}}
\def \ea{\end{array}}

\begin{document}
\title{High Mobility SiGe/Ge 2DHG Heterostructure Quantum Wells for Semiconductor Hole Spin Qubits}
\author{Zhenzhen Kong$^{1,2,3\dagger}$, Zonghu Li$^{4\dagger}$, Yuchen Zhou$^{4}$, Gang Cao$^{1,4,5*}$, Hai-Ou Li$^{1,4,5}$, Jiale Su$^{2}$, Yiwen Zhang$^{2,3}$,  Jinbiao Liu$^{2,3}$, Guo-Ping Guo$^{1,4,5}$, Junfeng Li$^{2}$,  Jun Luo$^{2,3}$, Chao Zhao$^{3}$, Tianchun Ye$^{3}$, and Guilei Wang$^{1*}$}

\email{wangguilei@hfnl.cn;gcao@ustc.edu.cn;}
\thanks{\\$\dagger$ These two authors contributed equally}

\affiliation{$^1$  Hefei National Laboratory, Hefei 230088, P. R. China}
\affiliation{$^2$  Integrated Circuit Advanced Process R\&D Center, Institute of Microelectronics, Chinese Academy of Sciences, Beijing 100029, P. R. China}
\affiliation{$^3$  School of Integrated Circuits, University of Chinese Academy of Sciences, Beijing 100049, P. R. China}
\affiliation{$^4$  CAS Key Laboratory of Quantum Information, University of Science and Technology of China, Hefei 230026, P. R. China}
\affiliation{$^5$  Origin Quantum Computing Company Limited, Hefei 230026, P. R. China}
\date{\today}

\begin{abstract}
Strong spin-orbit coupling and relatively weak hyperfine interactions make germanium hole spin qubits a promising candidate for semiconductor quantum processors. The two-dimensional hole gas structure of strained Ge quantum wells serves as the primary material platform for spin hole qubits. A low disorder material environment is essential for this process. In this work, we fabricated a Ge/SiGe heterojunction with a 60 nm buried quantum well layer on a Si substrate using reduced pressure chemical vapor deposition technology. At a temperature of 16 mK, when the carrier density is 1.87×10$^{11}$/cm$^{2}$, we obtained a mobility as high as 308.64×10$^{4}$cm$^{2}$/Vs. Concurrently, double quantum dot and planar germanium coupling with microwave cavities were also successfully achieved. This fully demonstrates that this structure can be used for the preparation of higher-performance hole spin qubits.
\end{abstract}

\maketitle

\section{Introduction}
In the realm of More than Moore and beyond CMOS, germanium has emerged as a promising material platform due to its superior properties in silicon-based optoelectronic devices and semiconductor quantum computing devices. Specifically, for the hole quantum spin bits, tunable g-factor and strong spin orbit interaction enable full electrical bit manipulation\cite{RN4,RN338,RN19}. Additionally, the reduced effective mass facilitates processing compared to the lithographic accommodation size of Si. The isotope $^{72}$Ge eliminates hyperfine interactions caused by nuclear spin similar with $^{28}$Si \cite{RN1072,RN13,RN370,RN986}. Currently, four spin qubits\cite{RN6, RN354} and a dozen quantum dots can be fabricated using two-dimensional holes in the strained Ge quantum well on the SiGe/Ge heterostructure, with the fidelity of hole spin qubits exceeding 99.9\%\cite{RN1167}. Furthermore, Ge materials hold significant application potential in physical systems such as superconductivity and topology. The strong Fermi-level pinning at the valence band of Ge materials enables good ohmic contact with most metals, significantly simplifying the process. Most importantly, the Si substrate allows direct compatibility with the ultra-large scale CMOS integration process. Thanks to the powerful micro-machining capability of this process, it holds significant importance for future multi-qubit expansion and chip miniaturization. \\
To effectively manipulate hole spin qubits, a material environment with low disorder and purity is essential. High mobility is a crucial evaluation criterion; in \textsc{iii-v}
materials a mobility exceeding 10 million mobility can be achieved\cite{RN1056}. However, the excessive nuclear spin  results in a  decoherence time on the nanosecond scale (for instance, the GaAs/AlGaAs decoherence time is  approximately 60 ns, while InSb and InAs have a decoherence time of about 10 ns)\cite{RN986}. While modulated doping can yield mobility excess of millions, the noise generated by dopant has shifted focus towards undoped SiGe/Ge heterostructure. In this SiGe/Ge structure,, the charge distribution transitions from equilibrium to non-equilibrium when the top barrier is approximately 50nm.\cite{RN37}. The thinner the top barrier, the more challenging it becomes to achieve high mobility due to the influence of remote impurity noise. Therefore, to obtain a high mobility pair so the material quality of the shallow buried layer must be superior. Currently, it is generally accepted that a 30-60 nm thick top barrier can effectively control the gate’s  influence on the charge in the quantum well, thereby effective qubit control \cite{RN1169}.\\
In this research, a SiGe/Ge heterostructure with a top SiGe barrier thickness of 60 nm was epitaxially grown on an 8-inch silicon substrate using RPCVD technology. This builds upon previous work that involved a shallow buried layer of 30 nm top barrier SiGe (ref\cite{RN1166}). An in-plane parallel compressive strain of $\varepsilon$\textsubscript{$\parallel$}
= -0.44\% was achieved at the Ge quantum well. The H-FET device then prepared, yielding an ultra-high hole mobility of $\mu$=308.64×10$^{4}$  cm$^{2}$/Vs under a charge density of 1.87$\times$10$^{11}$/cm$^{2}$ at 16 mK. Concurrently, signs of fractional quantum states were successfully detected. Double quantum dots were also prepared and their stable charge steady-state diagram was measured. Furthermore, the  coupling of hole quantum dot to microwave cavity in planar germanium was also successfully manufactured based on the SiGe/Ge heterostructure discussed in this article(ref\cite{RN240}). This fully demonstrates the low disorder of the material environment.

\section{MATERIALS AND METHODS}
In this letter, the SiGe/Ge heterostructure was fabricated using industrial RPCVD on a 200 mm standard Si p $\textless$
100$\textgreater$
 substrate preparation. 
A DHF (HF: H$_2$O 100:1) solution was used to eliminate the natural oxide layer on the wafer surface, Concurrently, a H suspension bond was generated on the cleaned surface which inhibited the formation of oxide film. 
Before epitaxy, the wafer surface was baked at a high temperature of H$_2$ exceeding 1000 $^\circ$C
 to remove the natural oxides on the Si substrate surface. DCS and GeH4 served as the precursors for epitaxial growth of SiGe layer, while germanane (GeH4) was utilized as the precursor for the Ge layer. Hydrogen was used as the carrier gas for this precursor, with a maintained H$_2$ flow rate of 20 slm. The structural design is depicted in Figure 1, wherein the strain quantum well thickness of the sample is 15 nm, the top barrier thickness is 60 nm, and the sample is capped by 1.4nm Si. The remainder of the structure is designed in accordance with ref \cite{RN1166}, incorporating a small tensile strained Ge buffer, a Ge reverse-graded SiGe, 90\% SiGe, and 80\% SiGe bottom barrier. The material’s thicknesswas determined using scan transmission electron microscope (STEM) and high resolution transmission electron microscope (HRTEM), while its composition was analyzed through energy dispersive spectrum (EDS). Strain was examined using high-resolution reciprocal lattice maps (HRRLMs), and surface roughness was detected via atomic force microscopy (AFM). TSTEM Morier imaging and Geometric Phase Analysis (GPA) analysis has been used to analyze strain changes on both sides of Si$_{0.2}$Ge$_{0.8}$ and Si$_{0.1}$Ge$_{0.9}$ abrupt layer structures. Gatan Digital Micrograph\textsuperscript{\textregistered}
 software\textsuperscript{\textregistered}
 is an in-house code used to analyze the GPA. The configuration and fabrication process of the Hall device were also consistent with ref\cite{RN1166}.

\section{RESULTS AND DISCUSSION}

\begin{figure}
\centerline{\includegraphics[width=9cm]{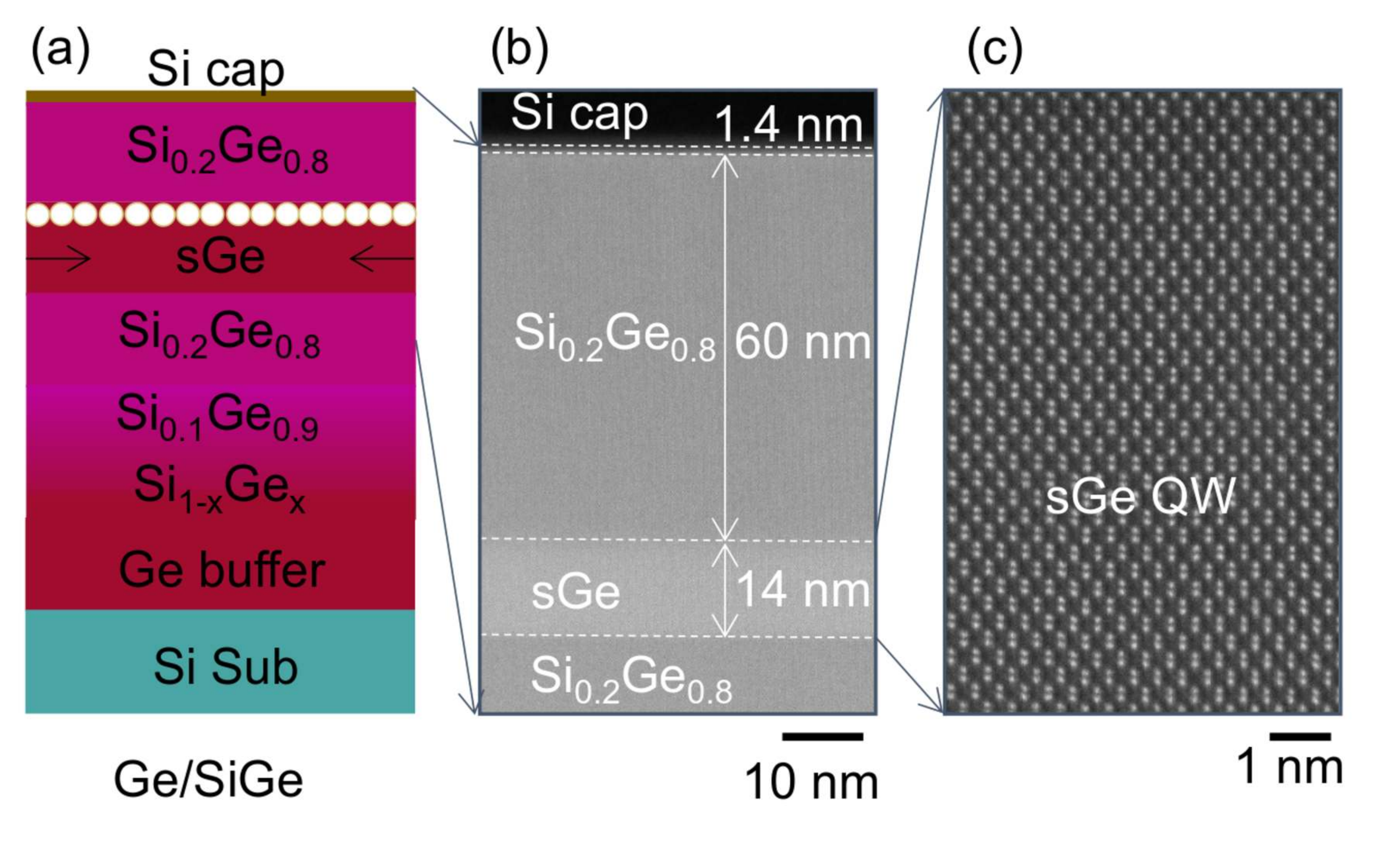}}
\caption{(a) Schematic of a Ge/SiGe heterostructure. Arrows represent the direction of compressive strain and hollow circles represent hole carriers. (b) TEM of 15 nm strained Ge quantum well and 60 nm top barrier SiGe. (c) Spherical Aberration Corrected Transmission Electron Microscope (AC-TEM) of Ge quantum well.}
\label{1}
\end{figure}

Schematic of a Ge/SiGe heterostructure is depicted in Figure~\ref{1}(a).From bottom to top , comprises a 1.7 $\mu$m Ge Virtual substrate, a 750 nm reverse grading layer with varying Ge contents transitioning from pure Ge to 90\% SiGe, a 200 nm Si$_{0.1}$Ge$_{0.9}$ constant component layer grown at 800 $^\circ$C, and a 360 nm Si$_{0.2}$Ge$_{0.8}$ bottom barrier grown at 650 $^\circ$C\cite{RN1166}. This abrupt interface reduced the majority of threading dislocation density being annihilated at this position. Subsequent analysis employs STEM Morier imaging and GPA analysis techniques as detailedin later in this article. A 15 nm compressive-strained Ge quantum well,a 60 nm top Si$_{0.2}$Ge$_{0.8}$ barrier, and a 1.4 nm Si cap are grown at 500 $^\circ$C, as illustrated in Figure~\ref{1} (b). The Si cap facilitates the formation of optimal interface states during device fabrication. The spherical high-resolution transmission microscope scanned the sGe QW, as shown in Figure~\ref{1} (c). High differential electron microscopy reveals a highly regular atomic arrangement within Ge quantum well. The design of the 60 nm top Si$_{0.2}$Ge$_{0.8}$ barrier is based on findings from reference 14, which identified impurity scattering at the semiconductor and gate medium as the primary source of transport test noise. Moreover, the top barrier of 60 nm does not compromise the effective manipulation of qubits.\\

\begin{figure*}
\centerline{\includegraphics[width=16cm]{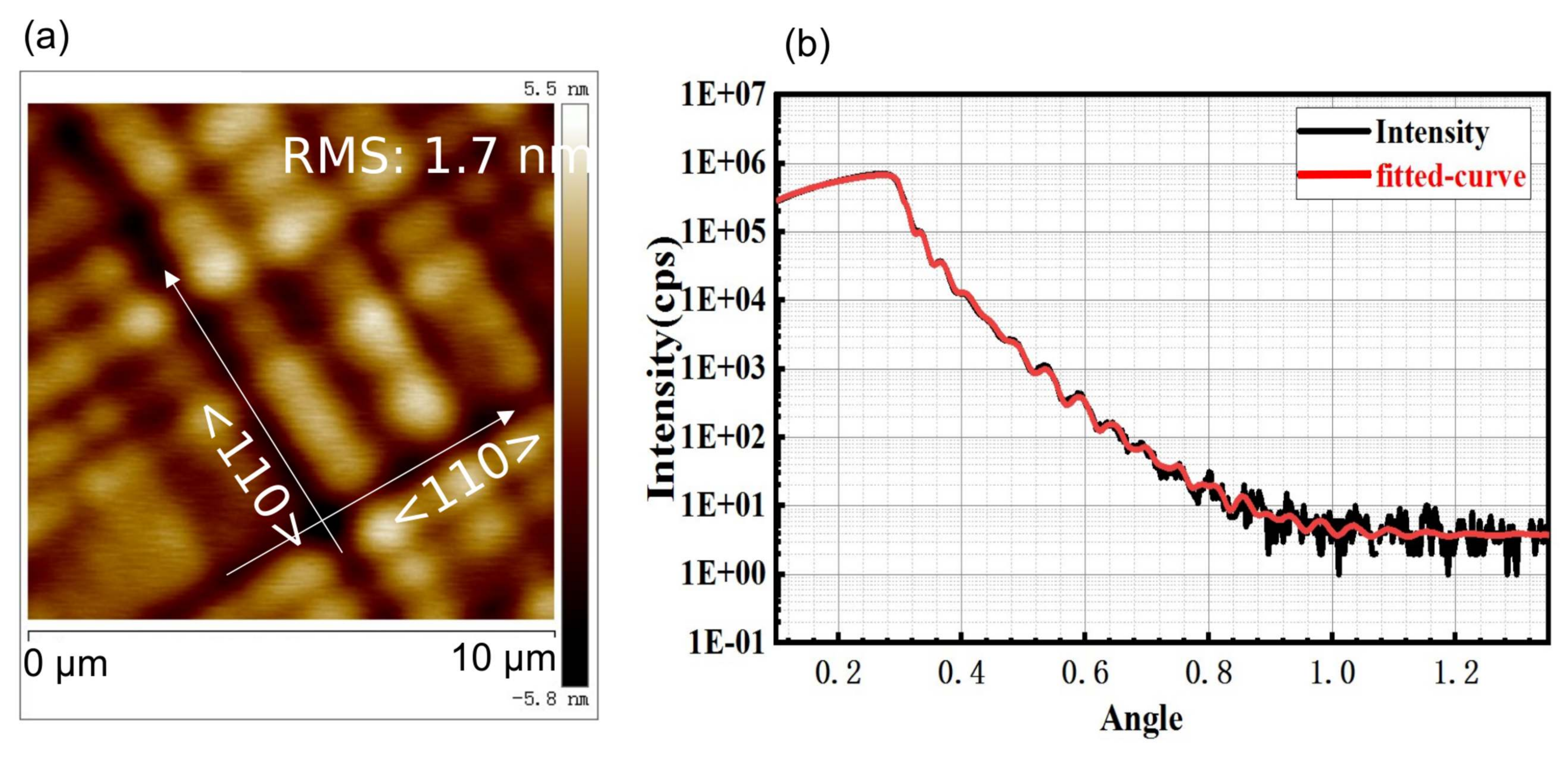}}
\caption{(a) AFM of the surface of Ge/SiGe heterostructure. The arrow represents $<$110$>$ Crystal phase. (b) XRR of Ge/SiGe 2DHG heterostructure.}
\label{2}
\end{figure*}

Figure~\ref{2}. (a) presents the surface of Ge/SiGe heterostructure scanned by AFM. The scanning area measures 10$\mu$m×10$\mu$m, and cross-hatch formed by strain release in the direction of $<$110$>$; The crystal phase aligns with the same direction as our Hall device and quantum dot device. The surface roughness is 1.7 nm, which is basically consistent with the 1.6099 nm calculated through X-ray Reflectivity (XRR) fitting.\\

\begin{figure}
\centerline{\includegraphics[width=9cm]{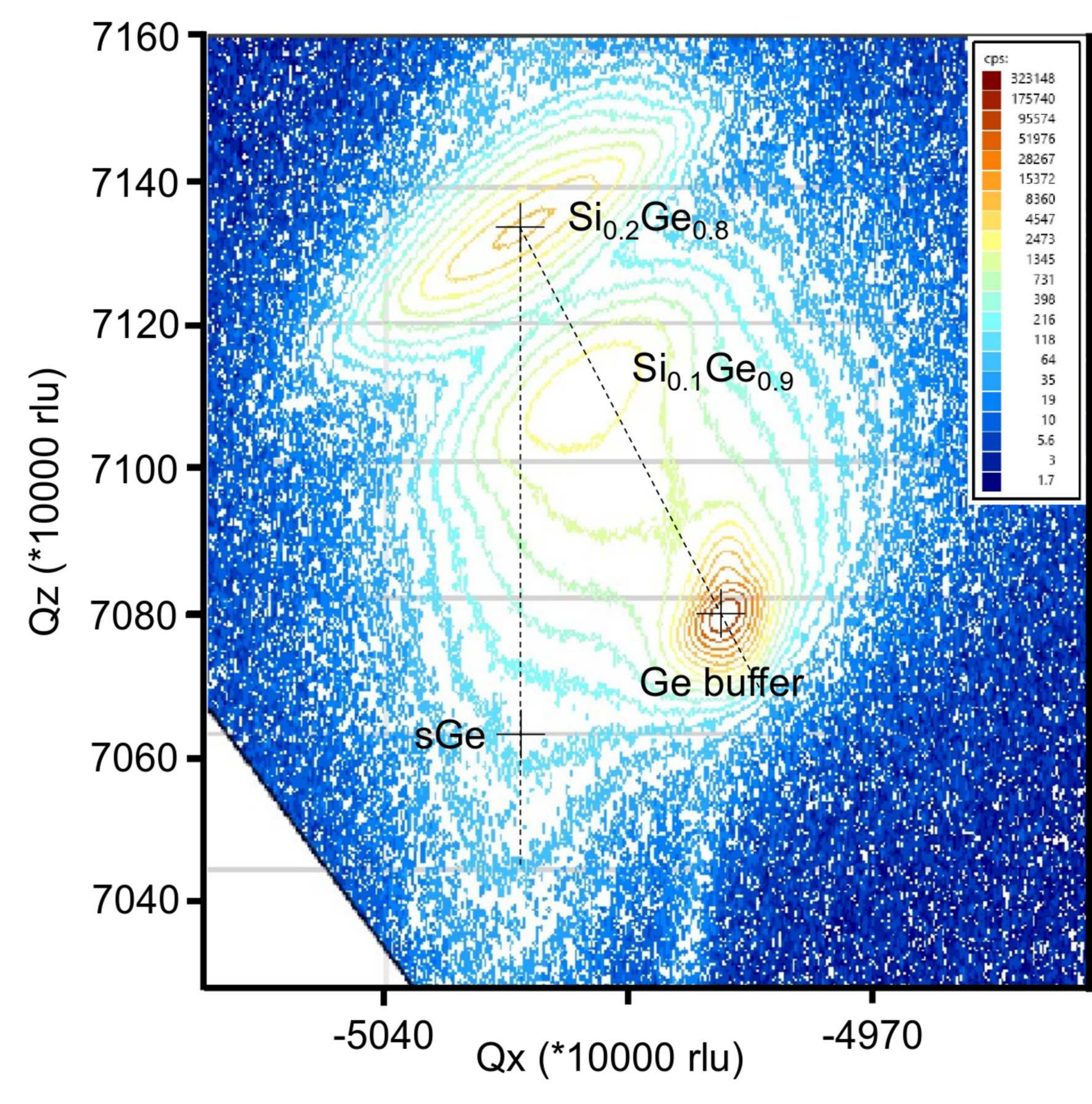}}
\caption{HRRLMs around (224) reflections.}
\label{3}
\end{figure}

To investigate the strain distribution within the material, we used HRRLMs to measure both the macroscopic strain and the corresponding crystal transformation in Ge quantum Wells. The HRRLMs were conducted using 224 asymmetric scans, as shown in Figure~\ref{3}.  It is important to note that the peak associated with the Si is not included in this Figure. The dotted line connecting SiGe barrier layer and Ge buffer layer is relaxed in accordance with the Si substrate. Both the sGe QW and the SiGe barrier layer are in a state of complete strain. The strain of sGe QW calculated by HRRLMs is -0.44\%.\\

\begin{figure*}
\centerline{\includegraphics[width=16cm]{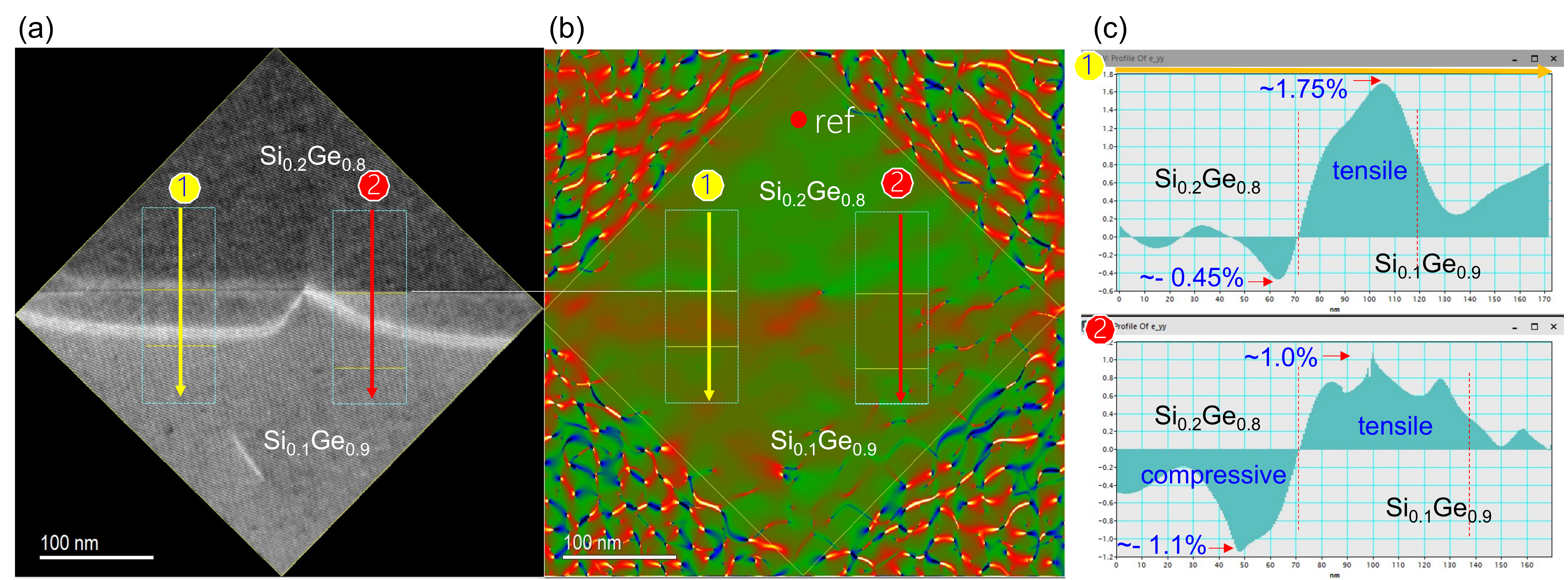}}
\caption{ (a) STEM Morier imaging and (b) GPA analysis. (c)Gatan Digital Micrograph\textsuperscript{\textregistered}
 software\textsuperscript{\textregistered}
 analysis of GPA}
\label{4}
\end{figure*}

At the Si$_{0.2}$Ge$_{0.8}$ and Si$_{0.1}$Ge$_{0.9}$ interface, we observed distinct dislocations that terminated at the interface, as shown in Figure~\ref{4}. (a). The Gatan Digital Micrograph\textsuperscript{\textregistered}
 software\textsuperscript{\textregistered}
 was utilized to analyze these defects as illustrated in Figure~\ref{4}. (b). This software Gatan digital micrograph was used to determine its strain, with the resultant stress distribution presented in Figure~\ref{4}. (c). Two different stress distributions are taken from the blue box in figure1 (a). The selected chosen region encompasses the Si$_{0.2}$Ge$_{0.8}$/Si$_{0.1}$Ge$_{0.9}$ interface, revealing that the dislocation clearly terminates at this interface. The red box in Si$_{0.2}$Ge$_{0.8}$ was adopted as a baseline to analyze the stress variations on the solid lines of yellow and red arrows. The stress under consideration is to the relative Si$_{0.2}$Ge$_{0.8}$ baseline, not reflecting on the actual relaxed material. The aim is to examine the impact of compositional and temperature variations at the interface on the overall stress distribution within the material. As depicted in Figure~\ref{4}. (c), a strain transformation is evident, suggesting that apart from inhibiting upward extension of the dislocations on either side of the interface, there is a significant strain release occurs. This phenomenon is likely associated with the rapid temperature fluctuation during growth, specifically from 800 $^\circ$C
 to 500 $^\circ$C.\\
.

\begin{table*}[htbp]
  \centering
  \caption{Summary of structure, mobility, and other parameters in different Ge/SiGe systems}
    \begin{tabular}{lrrrrr}
    \hline
     & \multicolumn{5}{c}{Key Parameters} \\
     \cline{2-6} 
    Study & \multicolumn{1}{c}{2DHG Depth (nm)} & \multicolumn{1}{c}{Ge content (\%)} & \multicolumn{1}{c}{Effective Mass ($m^*$)} & \multicolumn{1}{c}{Mobility ($\times 10^6$ cm$^2$/V.s)} & \multicolumn{1}{c}{Substrate/Size} \\
    \hline
    \textbf{[this work]} & \textbf{60} & \textbf{80\%} & \textbf{0.0787$m_0$} & \textbf{3.08} & \textbf{Si/200 mm} \\
    \textbf{ref \cite{RN1166} [our group]} & \textbf{32} & \textbf{80\%} & \textbf{0.0728$m_0$} & \textbf{2.4} & \textbf{Si/200 mm} \\
    \textbf{ref \cite{RN1170}} & \textbf{55} & \textbf{80\%} & \textbf{--} & \textbf{3.4} & \textbf{Ge/100 mm} \\
    \textbf{ref \cite{RN1168}} & \textbf{100} & \textbf{85\%} & \textbf{0.054$m_0$} & \textbf{4.3} & \textbf{Si/150 mm} \\
    \textbf{ref \cite{RN71}} & \textbf{66} & \textbf{90\%} & \textbf{0.068$\pm$0.001$m_0$} & \textbf{$\approx$} & \textbf{Si/100 mm} \\
    \textbf{ref \cite{RN19}} & \textbf{22} & \textbf{80\%} & \textbf{0.09$m_0$} & \textbf{0.55} & \textbf{Si/100 mm} \\
    \hline
    \end{tabular}%
  \label{table2}%
\end{table*}

\begin{figure}
\centerline{\includegraphics[width=9cm]{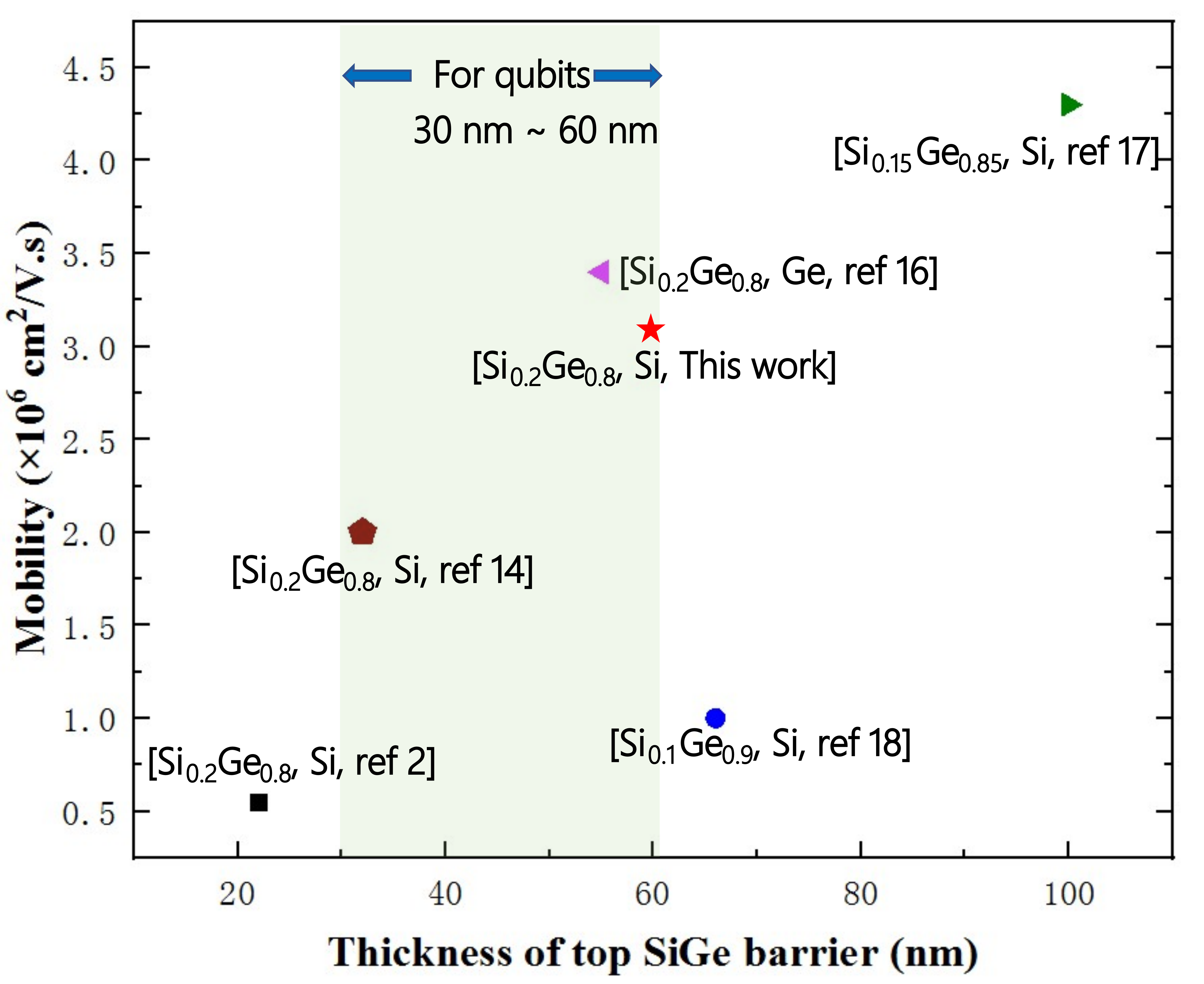}}
\caption{Research progress of low temperature mobility of undoped s-Ge QW by RPCVD in different research groups.}
\label{5}
\end{figure}

Table 1 summarizes the basic structure of undoped 2D hole gas, including barrier thickness and Ge components. At the same time, the carrier mobility, hole effective mass and substrate used under low temperature transport are summarized. The red star in Fig.~\ref{5}. and ref 14 denote the work of our research group. The horizontal axis signifies the thickness of the top SiGe barrier layer, with a current consensus suggesting that a thickness ranging from 30-60 nm is optimal for qubit manipulation. The square brackets indicate the content Ge component within both the SiGe barrier and the substrate utilized by the SiGe/Ge heterojunction.\\

\begin{figure*}
\centerline{\includegraphics[width=16cm]{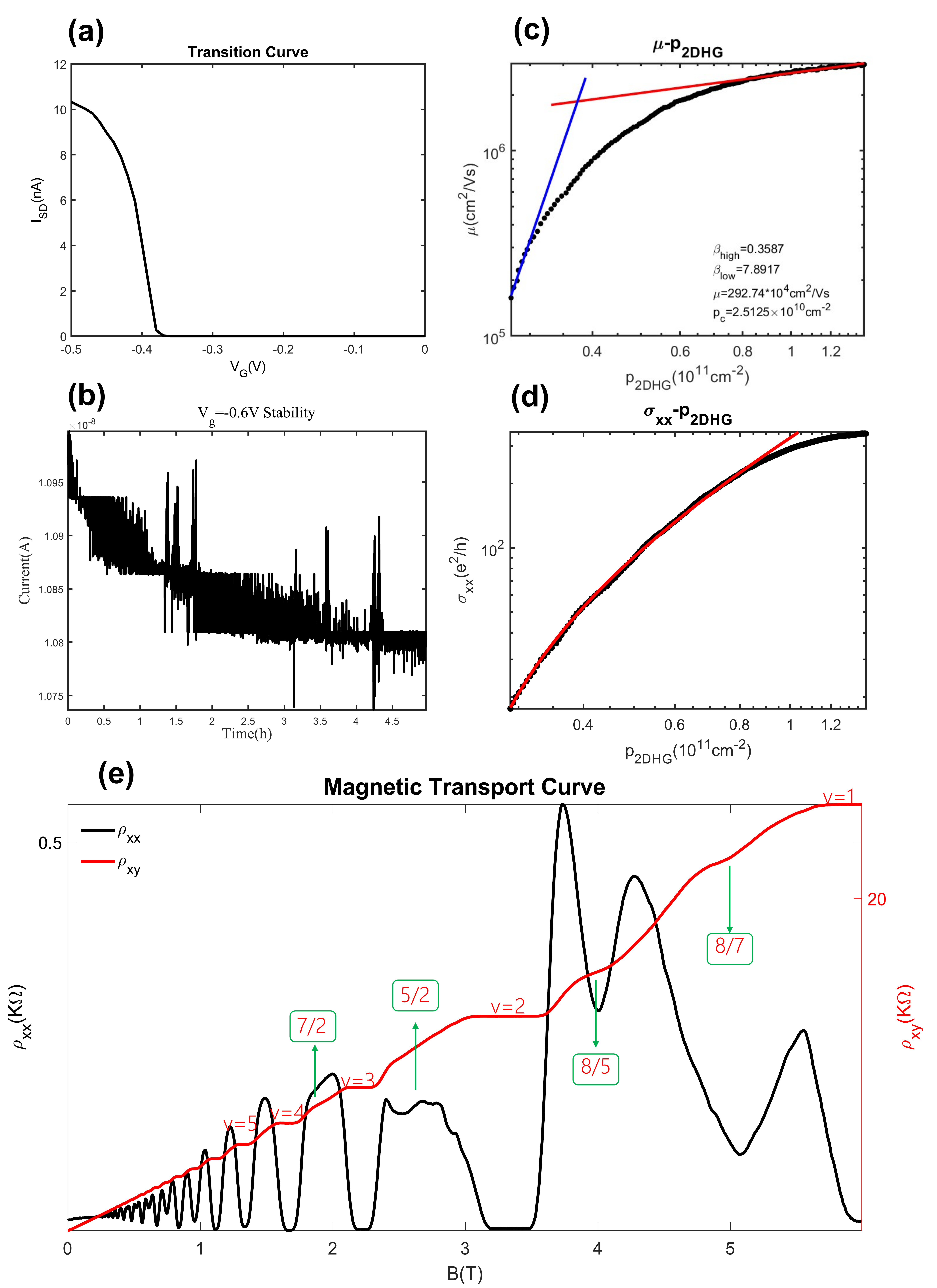}}
\caption{(a) Transfer characteristic curve. (b) Time stability testing of devices. (c) Mobility $\mu$ vs carrier density $p_{\text{2DHG}}$. (d) Longitudinal conductivity $\sigma_{xx}$ vs carrier density $p_{\text{2DHG}}$. (e) Magnetic transport curve at $(1.607 \pm 0.0473) \times 10^{11}/\text{cm}^2$.}
\label{6}
\end{figure*}

To further elucidate the material quality, Hall-bar devices were fabricated to characterize the transport properties of 2DHG heterojunctions. The detailed fabrication procedure for these Hall-bar devices can be found in Ref 14. The device's channel direction aligns with the $<110>$ in-plane crystallographic orientation of the material. Magnetic transport performance was assessed using a dilution refrigerator at a temperature of 16mK and a magnetic field strength of 6 T. Fig.~\ref{6}(a) presents the transmission characteristic curve of the device, with a scanning range from -0.6V to -0.01V (in steps of 0.01 V). From this figure, it is evident that the device's threshold voltage, is $V_{th} = (-0.36 \pm 0.01)\,V$. With a fixed voltage set at $V_g = -0.6V$. The device's stability over time was confirmed through continuous scanning, as depicted in Figure 6(b). After a duration of 5 hours, the device reached stability, with only 150 pA (1.36\%) reduction in current. The observed current jitter is approximately 54 pA. Notably, larger fluctuations are contributed to poor grounding conditions. Fig.~\ref{6} (c) illustrates the relationship between two-dimensional hole carrier density 2DHG and mobility $\mu$. The maximum mobility $\mu_{max}$ is $(292.7 \pm 0.9) \times 10^4\,\text{cm}^2/\text{Vs}$, and the peak carrier concentration determined through two-dimensional scanning is $p_{max} = (1.400 \pm 0.047) \times 10^{11}/\text{cm}^2$. We identified the high-density region and the low-density region using power law index fitting ($\mu = \beta p_{2DHG}^\gamma$). In the high-density region $\beta_{high} = (0.3587 \pm 0.0143)$, the scattering source is mainly background scattering. This value differs from the $\beta_{high} = 1.6968$ reported in a previous article by our research group (ref14). We posit that the original record includes impurity scattering in the high-density region obscured by the increased thickness of the SiGe top barrier, which has grown from 32 nm to 60 nm. This increase also contributes to the enhanced mobility of the 2DHG structure during this period, confirming that the mutation interface introduced beneath the bottom barrier effectively inhibits the penetration dislocation's extension into the quantum well. 

Through linear fitting, the critical density $p_c = 2.5125 \times 10^{10}\,\text{cm}^{-2}$ is obtained by calculating $\mu - \ln(p_{2DHG})$. According to the percolation theory, $\sigma_{xx} \propto (p_{2DHG} - p_p)^{2/3}$, we determined infiltration density $p_p = (2.045 \pm 0.037) \times 10^{10}/\text{cm}^2$, as depicted in Fig.~\ref{6} (d). There is also a further reduction compared to ref14, which indicates a further increase in the disorder of the material. Fig.~\ref{6} (e) presents the magnetic transport curve with a carrier concentration of $(1.607 \pm 0.0473) \times 10^{11}/\text{cm}^2$. Besides the prominent integer filling factor, there are discernible fractional quantum indications at positions such as 5/2, 8/5, and 8/7. The emergence of these fractional states signifies the interaction between charges in the high-purity material environment, thereby indicating the superior quality of the material. We conducted tests on various hall devices, at a carrier density of 1.87×10$^{11}$/cm$^2$, which yielded a maximum mobility density of $\mu = 308.64 \times 10^4\,\text{cm}^2/\text{Vs}$. At a carrier density of $(1.26 \pm 0.03) \times 10^{11}/\text{cm}^2$, the effective mass of holes $m^*$ can be obtained as $0.0787 m_0$ by fitting the temperature-damped decay. The effective $g^*$ factor is 10.816, indicating a certain degree of heavy and light hole mixing. Based on the provided material, the planar germanium quantum dots have been successfully coupled with the microwave cavity. This achievement establishes a solid foundation for the efficient manipulation of spin hole qubits.\\

\begin{figure*}  
\centerline{\includegraphics[width=13cm]{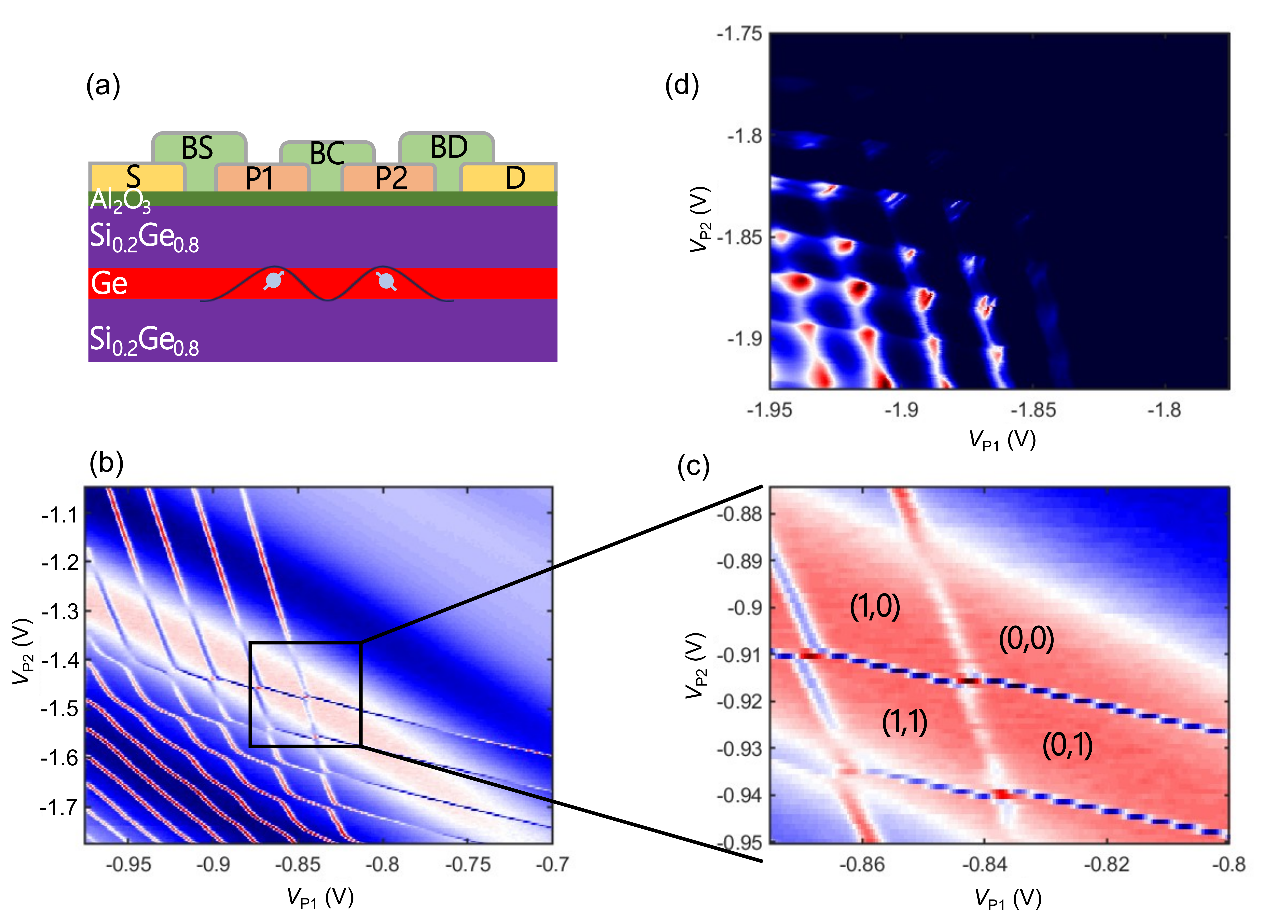}}
\caption{(a) Cross-section schematic of the quantum dot device. (b) Charge stability diagram of double quantum dot formed by sweeping voltage of plunger gates P1 and P2. (c) Close-up of the stability diagram in the few-electron regime. (d) Stability diagram of the double quantum dot.}
\label{7}
\end{figure*}

To assess potential of materials in the qubit preparation, we fabricated quantum dot devices. For a comprehensive description of the fabrication process, refer to Ref 6. Fig.~\ref{7} (a) presents the cross-section diagram of strained germanium double quantum dot devices. Fig.~\ref{7} (b) show differential conductance signals populated by hole states of double quantum dots, obtained through detection at a test temperature 20 mK. Each bright line signifies a hole tunneling event within double quantum dot, while parallel lines represent changes in hole filling for a single quantum dot. The region devoid of tunneling lines in the upper right corner signifies that the holes in the quantum have been depleted. The double quantum dots in the multi-hole region located in the lower right corner are progressively becoming single quantized due to capacitive coupling between quantum dots. The numbers (N1, N2) in Fig.~\ref{7} (c) denote the hole filling numbers of the left and right quantum dots respectively. By manipulating the source drain bias and top gate voltages VP1 and VP2, it is confirmed that double quantum dots can be defined and adjusted, as illustrated in Fig.~\ref{7} (d). This figure reveals a distinct representative bias triangle of the double quantum dots. This bias triangle serves as an effective characterization of Pauli spin blocking. As observed from the figure, the double quantum dot stability diagram is highly clear with no stray points, indicating that the strain-Ge quantum well is in a low-disorder state. This makes it suitable for preparing high-performance, high-fidelity spin hole qubits.

\section{Conclusions}
In summary, we have fabricated a high quality undoped Ge/SiGe heterojunction with an ultrahigh mobility over three million. Through comprehensive material characterization and low-temperature transport tests, it has been demonstrated that strained the germanium system has exhibits very low defects thereby providing a low-disorder environment conducive to quantum computing. The 60 nm top barrier effectively shields the distant impurity scattering at the gate dielectric interface, thus facilitating the effective manipulation of spin hole qubits. This is further corroborated by the successful preparation of double quantum dots and the successful coupling of planar germanium with microwave cavities.


\section*{Acknowledgements}

The authors declare no conflict of interest.
\bibliographystyle{apsrev4-1}
\bibliography{wenxianku}

\end{document}